\documentclass[11pt]{article}
\usepackage[dvips]{graphicx}
\usepackage[utf8]{inputenc}
\usepackage[english]{babel}
\usepackage[T1]{fontenc}

\usepackage[title,titletoc]{appendix}

\usepackage{graphicx} 
\usepackage{color} 
\usepackage{float}
\usepackage{tabularx}
\usepackage{indentfirst}
\usepackage{url}
\usepackage[hidelinks]{hyperref}

\newcommand{\ISSNno}{0103-9741}
\newcommand{\MCCSeqAno}{05/2019}
\newcommand{\TituloCapa}{A Ginga-enabled Digital Radio Mondiale Broadcasting Chain:
  Signaling and Definitions}

\newcommand{\AutorANome}{Rafael Diniz}
\newcommand{\AutorAemail}{rafaeldiniz@telemidia.puc-rio.br}
\newcommand{\AutorBNome}{Álan L. V. Guedes}
\newcommand{\AutorBemail}{alan@telemidia.puc-rio.br}
\newcommand{\AutorCNome}{Sergio Colcher}
\newcommand{\AutorCemail}{colcher@inf.puc-rio.br}

\usepackage{graphicx}
\usepackage{fancybox}
\usepackage{setspace}
\newenvironment{capalayout}{
    \setlength{\topmargin}{-3.0cm}
    \setlength{\footskip}{0cm}
    \thispagestyle{empty}
} 

\begin{document}

\begin{capalayout}
        \centering
        \includegraphics[keepaspectratio,width=14.7cm]{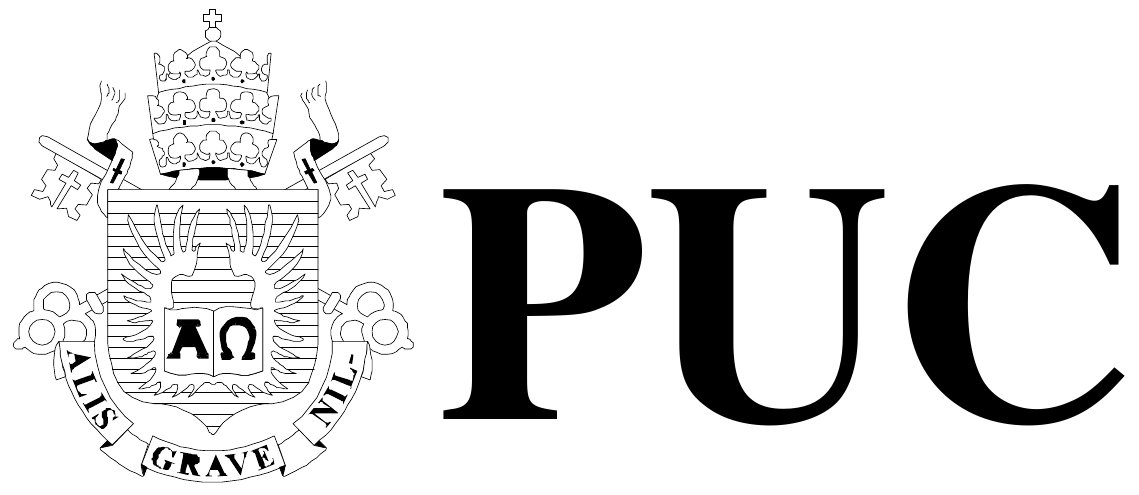}
        
        \medskip
    
        \setlength\fboxsep{1pt}
        \shadowbox{\fbox{\begin{minipage}[h]{14.5cm}
            \begin{center}
                \doublespacing
                \fontfamily{phv} \fontsize{14}{16} \selectfont
                \medskip
                ISSN \ISSNno
                
                \bigskip
                Monografias em Ciência da Computação 
                
                n\textordmasculine \, \MCCSeqAno
                
                \bigskip
                \medskip
                \fontsize{18}{20}\selectfont
                \textbf{\TituloCapa}
                
                \bigskip
                \fontsize{14}{15}\selectfont
                \textbf{\AutorANome} \\
                \textbf{\AutorBNome} \\
                \textbf{\AutorCNome}
                
                \bigskip
                
                \medskip
                Departamento de Informática
                \bigskip
            \end{center}
        \end{minipage}}}
        
        \bigskip
        \bigskip
        
        \begin{minipage}[h]{14.5cm}
            \doublespacing
            \fontfamily{phv}
            \begin{center} 
                \fontsize{12}{14}\selectfont
                \textbf{PONTIFÍCIA UNIVERSIDADE CATÓLICA DO RIO DE JANEIRO} \\
                \textbf{RUA MARQUÊS DE SÃO VICENTE, 225 - CEP 22451-900} \\
                \textbf{RIO DE JANEIRO - BRASIL}
            \end{center}
        \end{minipage}
        \newpage
    \end{capalayout}

\thispagestyle{empty}

\begin{flushleft}
\begin{tabular}{p{11.1cm}r}
Monografias em Ciência da Computação, No. \MCCSeqAno & ISSN: \ISSNno \\
Editor: Editor: Prof. Carlos José Pereira de Lucena   & Junho, 2019
\end{tabular}
\end{flushleft}
\LARGE
\bigskip
\begin{center}
    {\bf \TituloCapa}
\end{center}
\normalsize
\bigskip
\begin{center}
{\bf \AutorANome, \AutorBNome~ and \AutorCNome}
\end{center}
\begin{center}
   \AutorAemail, \AutorBemail, \AutorCemail
\end{center}

\noindent {\bf Abstract.} ISDB-T International standard is currently adopted by most Latin America
    countries. To support interactive applications in Digital TV receivers, ISDB-T
    defines the middleware Ginga. Similar to Digital TV, Digital Radio standards
    also provide the means to carry interactive applications; however, their
    specifications for interactive applications are usually more restricted than
    the ones used in Digital TV.
    Also, interactive applications for Digital TV and Digital Radio are usually
    incompatible.
    Motivated by such observations, this report considers the importance of
    interactive applications for both TV and Radio Broadcasting and the advantages
    of using the same middleware and languages specification for Digital TV and
    Radio.
    More specifically, it establishes the signaling and definitions on how to
    transport and execute Ginga-NCL and Ginga-HTML5 applications
    over DRM~(Digital Radio Mondiale) transmission.
    Ministry of Science, Technology, Innovation and Communication of Brazil is
    carrying trials with Digital Radio Mondiale standard in order to define the
    reference model of the Brazilian Digital Radio System (Portuguese: Sistema
    Brasileiro de Rádio Digital - SBRD).
\medskip

\medskip

\noindent {\bf Keywords:} Digital Radio; Digital Radio Mondiale; Middleware; Ginga; NCL \\

\bigskip
\noindent {\bf Resumo.} O padrão internacional ISDB-T é atualmente adotado pela maioria
dos países
da América Latina. Para suportar aplicações interativas em receptores de
TV digital, o ISDB-T
define o middleware Ginga. Similar à TV Digital, padrões de Rádio Digital
também fornecem os meios para transportar aplicativos interativos; no
entanto, suas especificações
para aplicações interativas são geralmente mais restritas da que as
usadas na TV
Digital. Além disso, aplicações interativas para TV Digital e Rádio
Digital são geralmente
incompatíveis. Motivado por tais observações, este relatório considera a
importância de
aplicações interativas para TV e Radiodifusão e as vantagens de usar a
mesma especificação
de middleware e linguagens para TV Digital e Rádio Digital. Mais
especificamente, estabelece
a sinalização e as definições de transporte e execução de aplicativos
Ginga-NCL e
Ginga-HTML5 sobre transmissão DRM (Digital Radio Mondiale). O Ministério
da Ciência,
Tecnologia, Inovações e Comunicações do Brasil está realizando ensaios
com a norma Digital
Radio Mondiale, a fim de definir a modelo de referência do Sistema
Brasileiro de Rádio
Digital (SBRD).

\medskip
\medskip

\noindent {\bf Palavras-chave:} Rádio digital; Rádio Digital Mundial; Middleware; Ginga; NCL \\

\newpage
\pagenumbering{roman} \setcounter{page}{2}
\vspace*{\fill}
\begin{flushleft}
    \textbf{In charge of publications:} \\
    PUC-Rio Departamento de Informática - Publicações \\
    Rua Marquês de São Vicente, 225 - Gávea \\
    22453-900 Rio de Janeiro RJ Brasil \\
    Tel. +55 21 3527-1516 Fax: +55 21 3527-1530 \\
    E-mail: publicar@inf.puc-rio.br \\
    Web site: http://bib-di.inf.puc-rio.br/techreports/ \\
    \end{flushleft}

\newpage
\renewcommand{\contentsname}{Table of Contents }
\renewcommand{\appendixname}{Annex}
\tableofcontents

\newpage
\pagenumbering{arabic} \setcounter{page}{1}

\section{Introduction}
In a Digital Television System, the middleware is the software layer that is
part of the receiver and is responsible for receiving and executing
(interactive)~applications that can be sent by the broadcaster. It must be
standardized by the Digital TV System~\cite{morris_interactive_2005} so
that any receiver, from any manufacturer, is able to receive and interpret
applications.

\begin{figure}[H]
  \centering
  \includegraphics[scale=0.5]{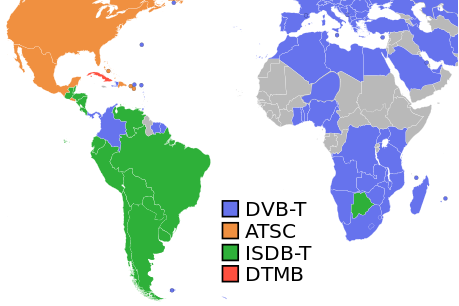}
  \caption{Countries painted in green are the ones where Ginga is used as DTV
           middleware.}
  \label{figureginga}
\end{figure}

The Brazilian Digital TV System\null~\footnote{In 2009, the Brazilian DTV System was rebranded as
``ISDB-T International'' and harmonized with the original ISDB-T. The main
differences between the Japanese ISDB-T and ISDB-T International are the
audio and video codecs, which were upgraded from MPEG~2 to MPEG~4, and the
application middleware, which is a BML~(Broadcast Markup Language)-based
technology, in case of the original ISDB-T, and Ginga, in case of ISDB-T
International. See
\emph{http://www.dibeg.org/techp/aribstd/harmonization.html} for more
information.
},
which is an evolution of the
Japanese ISDB-T, defines the middleware Ginga~\cite{Soares-L-F-G-2010b},
and its declarative language, NCL~(Nested Context
Language)~\cite{ABNT-15606-2-2015}, to support interactive applications.
It is currently adopted by 14 countries in Latin America (the ones in green
in Figure~\ref{figureginga}) and is an ITU-T recommendation for IPTV
systems~\cite{ITU-T-H761-2014}.
Indeed, by definition, Ginga is an IBB (Integrated Broadcast-Broadband) system,
as specified in ITU-R BT.2267-5~\cite{ITU-2267-5-2015}.
This means that the middleware permits that applications received via broadcast
channel ---e.g., broadcast radio applications--- use the Internet as a
return channel, whenever such support is available in the receiver.
Ginga applications are written in NCL (Nested Context Language), which is a
declarative, domain-specific language for the description of interactive
multimedia presentations. Recently (2018) the Brazilian DTV Forum
added HTML5 as a supported application to Ginga (together with
NCL), as specified by ABNT 15606-10\cite{ABNT-15606-10-2018} named
Ginga-HTML5, which is a strict subset of W3C HTML specification. Also
added to the middleware of the Brazilian DTV in 2018 was the Ginga Common
Core WebServices, which allows native applications of the digital receiver
ecosystem (eg. Android, WebOS, Tizen) to interoperate with Ginga
applications through a REST API, as defined by ABNT
15606-11\cite{ABNT-15606-11-2018}.

Like in DTV systems, Digital Radio must also permit the transmission of
applications. Radio broadcasters have very different coverage, content and
audience, ranging from high power public Broadcasters, which cover the whole
Amazon region, high power commercial broadcasters, educative and small
community radios. Applications will be able to target very different
contexts, be it providing public services, traffic information, games,
multimedia content related to the audio, advertisement and so on.

This report describes the transport of Ginga NCL and Ginga HTML applications in
the Digital Radio Mondiale~\footnote{DRM is a digital radio broadcasting system standardised
for all broadcasting frequencies} (DRM) system ---i.e., it gives a set of
definitions that allow for such transport--- and discusses minor adaptations
to Ginga and NCL that can improve their support to digital radio-specific
application requirements. The proposed adaptations are defined as amendments to
the 2014 ITU-T~H.761~\cite{ITU-T-H761-2014} text. Thus, this report may be used
as a reference for a possible standardization of the use of Ginga in the DRM
system. As a consequence, it is also a contribution to the Brazilian Digital
Radio System specification~\footnote {Brazilian
Digital Radio System specification, in Portuguese, Sistema Brasileiro de
Rádio Digital, was established in March 30, 2010, but has no reference model
defined until today. See:
http://www.abert.org.br/web/index.php/legistecnica/item/portaria-n-290-de-30-marco-de-2010}.

The remainder of the report is organized as follows.
Section~\ref{sec:transport} discusses proposed signalling extensions
to support the transmission and correct interpretation of NCL applications
using the DRM transmission structure.
Section~\ref{sec:ncl_extensions} proposes NCL extensions that are useful in
Digital Radio scenarios.
Section~\ref{sec:nclua_extensions} brings extensions to the NCLua
APIs~\footnote{NCLua is the scripting language supported by Ginga. NCLua
  scripts can be called a from a NCL application.}.
Section~\ref{sec:receiver_profile} presents receiver profiles that may be
useful in the context of Digital Radio broadcasting.
Finally, Section~\ref{sec:conclusion} discusses our main conclusions and future
work. An annex~\ref{sec:annex} also is present in the end, with complimentary information.

\section{On the transport of Ginga applications over DRM}
\label{sec:transport}

This section describes how to multiplex Ginga NCL and Ginga HTML applications over
DRM. The multiplex scheme used in DRM is described in the DRM System
Standard~\cite{ETSI-201980-2014}, where also the modulation and channel
coding, transmission structure and source coding are defined.

\subsection{MOT protocol}
The DRM multiplex defines a transmission scheme consisting of three logical channels: the
\emph{Main Service Channel}~(MSC), the \emph{Fast Access Channel}~(FAC),
and the \emph{Service Description Channel}~(SDC). MSC carries the data for all the DRM Streams in the DRM multiplex;
it may contain between one and four DRM Streams, and each DRM Stream may be
either audio or data (in this case, a DRM Stream can carry up to 4
sub-stream in Packet Mode).
FAC provides information on the channel width and related parameters, and it
also provides service selection information, allowing for fast scanning.
Finally, SDC gives information on how to decode the MSC and provides
descriptors defining the DRM Services within the multiplex.

As aforementioned, DRM defines two type of services (DRM Services): \emph{audio} and
\emph{data}.
An audio service must be associated with an audio DRM Stream and
can be optionally associated with data DRM Streams as \emph{Program
  Associated Data}~(PAD). A data service must be associated with a
data DRM Stream.

\begin{figure}[H]
  \centering
  \includegraphics[scale=0.3]{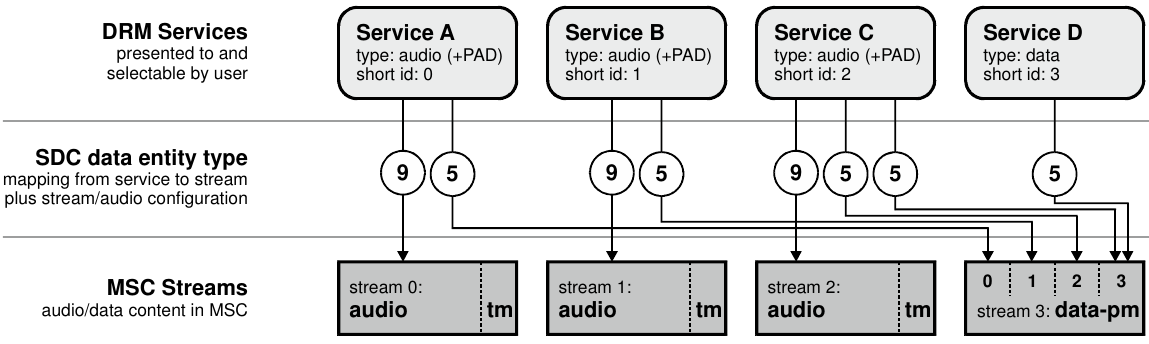}
  \caption{An example of a DRM multiplex configuration composed of three audio
           services~(A-C) using the Program Associated Data and one standalone
           data service~(D)}.
  \label{figuredrm}
\end{figure}

In the case of data DRM streams, DRM multiplex provides a~\emph{Packet
  mode}, which defines a generalized way to deliver packetized data. The
data stream can be associated by one or more \emph{standalone data service}
or \emph{associated to an audio service} as PAD. The mapping between the
DRM Services and the DRM Streams is defined by SDC entities.
Figure~\ref{figuredrm} shows an example of multiplex configuration composed of
four DRM Services.

Data streams can carry interactive applications, to be executed by the
receiver.

To signalize the transmission of Ginga applications as standalone data
services~\footnote{We
  assume that most applications will be transmitted as PAD of audio service,
  and not as standalone data service.}
in the DRM multiplex, Table~\ref{tablefac} proposes a FAC parameter
\emph{Application identifier} with value $4$ ---the first available
``reserved for future definition'' in ETSI TS~101~968~\cite{ETSI-101968-2009},
which is the standard that contains the identifiers of DRM applications.

\begin{table}[H]
\centering
\caption{FAC Application identifier value for Ginga application transmitted as
  standalone data service.}
\label{tablefac}
\begin{tabular}{|l|l|}
  \hline
  FAC Service Parameter & Value \\
  \hline
  Service Descriptor (Application identifier) & 4 \\
  \hline
\end{tabular}
\end{table}

To carry applications composed of a set of files, which is usually the
case of an NCL application, DRM uses the DAB MOT
protocol~\cite{ETSI-301234-2006}. MOT (Multimedia Object Transfer) is a
protocol that allows for the transmission of one or more files in a cyclic
way~(i.e., as a carousel).
It is already used in standardized digital radio applications such as
SlideShow~\cite{ETSI-101499-2015} and Broadcast
Website~\cite{ETSI-101498-1-2006}. Similarly, the files that are part of an
NCL application may also be carried as MOT objects.

MOT objects are segmented in DAB MSC data groups. Data groups are mapped
directly to DRM data units.
As detailed in Chapter~5.2 of ETSI TS~101~968~\cite{ETSI-101968-2009},
DRM data units are subsequently split into packets that are transported by
the DRM Packet mode protocol. To reference an NCL
application, the \emph{Application information} parameters, which
identifies an application in the SDC and associates the application to a
service, should be set to those presented in Table~\ref{tablesdc}.
The first two parameters, \emph{Packet Mode Indicator} and \emph{Data Unit
Indicator}, are required by the MOT protocol; an application domain with
value~$0$ indicates a DRM application, and the user application domain with the
proposed value~$0x0001$ indicates an NCL application.
Note that $0x0001$ is the first available application identifier value for
openly specified applications in ETSI~TS~101~968~\cite{ETSI-101968-2009}.

\begin{table}[H]
\centering
\caption{SDC Application information parameter values for NCL applications
  transmitted using MOT protocol.}
\label{tablesdc}
\begin{tabular}{|l|l|}
  \hline
  Application information parameters & Value \\
  \hline
  Packet mode indicator & 1 \\
  \hline
  data unit indicator & 1 \\
  \hline
  application domain & 0 \\
  \hline
  user application identifier & 0x0001 \\
  \hline
\end{tabular}
\end{table}

Since Ginga expects that the NCL application files are organized into a directory
tree, the MOT protocol's Directory Mode must be used. The
\emph{Directory Mode} provides support for transmitting many files organized
into a directory tree with the possibility of transmitting them in an
interleaved way. The mandatory \emph{DirectoryExtension} parameter is presented
in Table~\ref{dirext}.
The \emph{DirectoryExtension} parameter contains information that apply to the
whole MOT transmission.

\begin{table}[H]
\centering
\caption{Mandatory MOT DirectoryExtension parameter.}
\label{dirext}
\begin{tabular}{|l|l|}
  \hline
  Parameter Id & Parameter \\
  \hline
  0x22 (100010) & DirectoryIndex \\
  \hline
\end{tabular}
\end{table}

The syntax of \emph{DirectoryIndex}
parameter's data field is shown in Table~\ref{dirsyn}.

\begin{table}[H]
\centering
\caption{Syntax of DirectoryIndex parameter's data field.}
\label{dirsyn}
\begin{tabular}{|l|l|}
  \hline
  Syntax & Size\footnote{All types are Unsigned Integer, transmitted Most Significant Bit First} \\
  \hline
  DirectoryIndex\_parameter\_data\_field() \{ &  \\
  \hline
  \hspace{5mm} profile\_id & 8 bits \\
  \hline
  \hspace{5mm} for (i=0;i<N;i++) \{ & \\
  \hline
  \hspace{10mm} entry\_point\_byte & 8bits \\
  \hline
  \hspace{5mm} \} & \\
  \hline
  \} & \\
  \hline
\end{tabular}
\end{table}

The \emph{DirectoryIndex} parameter indicates the application entry
point~(\emph{entry\_point} field)
for a given receiver profile~(\emph{profile\_id} field).  One may insert
more than one \emph{DirectoryIndex} parameter in order to signalize
different entry points for distinct receiver profiles. The
\emph{entry\_point} is composed of \emph{entry\_point\_byte} fields, which
must be ISO/IEC~10646~\cite{ISO-10646-2014} characters (using UTF-8
transformation format), being the hash sign~(`\#') a
reserved character. The entry point must follow one of the possible syntaxes
expressed in Table~\ref{entrypt}.  In the first syntax, one specifies a NCL
or HTML application to be started, while in the second syntax, one specifies
both the NCL file and a specific interface~(\emph{<port>}) to be started.  Note that
in the second syntax the file name and the port identifier must be separated
by a hash sign~(`\#'). In both syntaxes the file name must be a relative
path, i.e., one not starting with character~`/', e.g., ``code/main.ncl'' or
``code/index.html''.

\begin{table}[H]
\centering
\caption{Entry point syntax.}
\label{entrypt}
\begin{tabularx}{\textwidth}{|p{7cm}|X|}
  \hline
  Entry point syntax & Description\\
  \hline
  \{application\_filename\}.\{ncl,html\} & The middleware should use the
  NCL or HTML application as entry point according to file name extension.\\
  \hline
  \{application\_filename\}.ncl\#\{InterfaceId\} & The middleware should use port
  \emph{InterfaceId} to start the NCL application \emph{application\_filename.ncl}.\\
  \hline
\end{tabularx}
\end{table}

Another optional \emph{DirectoryExtension} parameters that must be supported
by the Ginga middleware running in the receiver are
\emph{SortedHeaderInformation}, \emph{DefaultPermitOutdatedVersions}, and
\emph{DefaultExpiration}.  Their semantics is the same specified in the MOT
standard~\cite{ETSI-301234-2006}.

The MOT protocol contains parameters related to the individual files,
transmitted by the MOT structures.  All files transmitted over the MOT
have, among other parameters, two mandatory parameters with file type
information: \emph{ContentType} and \emph{ContentSubType}\null
\footnote{The possible values specified in Table~17 of ETSI
          TS~101~756~\cite{ETSI-101756-2015}}.
The Ginga middleware must ignore these values, as they do not specify all
supported file types supported by Ginga.
The recommended values for these fields are shown in Table~\ref{types}.

\begin{table}[H]
\centering
\caption{Recommended values for ContentType and ContentSubType.}
\label{types}
\begin{tabular}{|l|l|}
  \hline
  Field & Value \\
  \hline
  ContentType & 0 \\
  \hline
  ContentSubType & 0 \\
  \hline
\end{tabular}
\end{table}

Moreover, the MOT protocol defines optional parameters for each file. These
parameters are defined
in the \emph{Header extension} part of the protocol header and their
use is specified in Table~\ref{context}.

The \emph{ContentName} parameter specifies a relative path for each file
that composes an application (e.g., ``media/pic.jpg'') and the
\emph{CompressionType} specifies compression algorithm used to compress the
file, when this is transmitted compressed.

Other parameters can be optionally present in the Header extension.
The optional parameters that must be correctly interpreted by the Ginga
middleware are the following: \emph{PermitOutdatedVersions},
\emph{Expiration}, and \emph{TriggerTime}.  The semantics of these
parameters is specified in the MOT standard~\cite{ETSI-301234-2006}.

\begin{table}[H]
\centering
\caption{Header extension parameters.}
\label{context}
\begin{tabularx}{\textwidth}{|X|X|X|}
  \hline
  Identifier & Parameter & Content \\
  \hline
  0x0C (001100)
  & ContentName
  & Contains the character set indicator, which must be set to ISO/IEC~10646
  as specified in Table~19 of ETSI TS~101~756~\cite{ETSI-101756-2015}
  (value~1111b), and the file name of the content, which must use a relative
  path.  This parameter is mandatory.\\
  \hline
  0x11 (010001)
  & CompressionType
  & Must be used when a file is transmitted compressed.  The only allowed
  compression is GZip (value~0x01) as specified in Table~18 of ETSI
  TS~101~756~\cite{ETSI-101756-2015}.  The middleware must support GZip
  decompression.\\
  \hline
\end{tabularx}
\end{table}

\section{Extensions to NCL}
\label{sec:ncl_extensions}

The use of NCL in the radio context is similar to~(and most time compatible
with) its use in DTV or IPTV contexts. However, there are some differences that
are explained in this section. Motivated by these differences, the remainder of
this section propose an NCL profile to Digital Radio.

\subsection{The Digital Radio Profile}
The Digital Radio (DR) NCL 3.1 profile is based on the
ITU H.761 NCL 3.1 Enhanced DTV (EDTV) profile, with the following differences:

\begin{itemize}
\item \emph{Transition} and \emph{TransitionBase} modules of NCL are not
      defined in the DR profile. The <transition> and <transitionBase> elements
      are not defined, nor it is allowed to reference transition via <property>
      element;
\item \emph{clip} and \emph{coords} attributes of the <area> element are
      not defined.
\item the properties \emph{transIn}, \emph{transOut} and \emph{plane} are
      not defined to be used in the element <property>.
\item global variables in NCL are defined as special properties of a media with
      type equals to \emph{application/x-ncl-settings} type. The following
      variables of the system group are not defined: \emph{screenVideoSize},
      \emph{screenBackgroundSize}, \emph{screenGraphicSize} and
      \emph{screenGraphicSize(i)}. This is because different graphic planes
      are supported in DR profile.
\item In the \emph{si} group of local variables, new variables are introduced
      in the DR profile: \emph{stationLabel}, \emph{numberOfServices},
      \emph{channelFrequency}, \emph{signalQuality}, and
      \emph{serviceDecoding}. \emph{stationLabel} contains the Broadcaster
      label. \emph{numberOfServices} contains the number of services present
      in the received signal, signalQuality contains the Modulation
      Error Ratio (MER), in decibels (only positive values, 0 meaning no
      signal). And, \emph{serviceDecoding} is a boolean value, \emph{True}
      or \emph{False}; being \emph{True} when the Bitrate Error Rate (BER)
      is less then \(10^{-4}\), meaning Quasi Error Free (QEF) reception.
\item the metadata group of local variables are not defined.
\end{itemize}

\subsection{Extra URI definitions}
The \emph{dsm-cc:} and \emph{ts:} URI schemes are not supported in the Ginga
DR profile, and the scheme \emph{drm:} defined in
ETSI TS 103 270~\cite{ETSI-103270-2015}, Table 9 (DRM parameter description)
and Table 12 (Example of RadioDNS bearerURI construction for DRM) shall be
supported.

\section{Extensions to the NCLua API}
\label{sec:nclua_extensions}
NCLua is the set of APIs that supports the integration between NCL
documents and the Lua scripting language.
NCLua is standardized in~\cite{ITU-T-H761-2014,ABNT-15606-2-2015}, and provides
two main modules: \emph{canvas}, which supports 2D~drawing primitives; and
\emph{event}, which provides the event-based communication between Lua and
NCL, and general event handling.

The Lua version 5.3 must be be used to implement the NCLua player in the
Ginga middleware for radio.

\subsection{The \emph{geolocation} class}
Given the mobile nature of Digital Radio, location-aware applications play an
important role in this context.
Aware of such a feature, and due to the lacking of similar APIs in the DTV
contest, this report proposes to extend the NCLua event API with a
``geolocation'' class. Geolocation events carry the global position, speed, and
heading of the receiver---of course, if the receiver can resolve such queries.

An NCLua script requests the geolocation information of the receiver by posting
an event of the form:

\begin{verbatim}
evt = { class = 'geolocation', [timeout=number] }
\end{verbatim}

\emph{timeout} is an optional field that, if present, determines the
timeout~(in seconds) for the query to be successfully completed.

An NCLua script receives the answer of geo-location request as event of the
form:

\begin{verbatim}
evt = { class = 'geolocation',
        [latitude=number], [longitude=number],
        [accuracy=number], [speed=number], [heading=number] }
\end{verbatim}


\section{Ginga Full Receiver Profile}
\label{sec:receiver_profile}

Recent tabletop receivers, media centers, TV with radio tuner, infotainment
automotive receivers, and mobile phones with radio support can easily run
the Ginga middleware and media decoders.

At least a touch screen or keypad with directional keys plus ``ENTER'' key
is required as interface to the receiver. A ``BACK'' key is desired in a
receiver with keypad.

No inferior or superior screen resolution and size are defined. It's
recommended for the receivers to have at least a 320x240 pixels color
screen.

As the definition of receiver profiles depends on many industrial production
aspects, just one receiver profile will be defined, leaving room for future
definitions like supported resolutions for images and audio encoder
features. This sole profile is named ``Ginga Full Receiver Profile'', with
\emph{profile\_id} in Table~\ref{dirsyn} equal to 1. A ``Ginga Full Receiver
Profile'' compatible receiver shall be compatible with all the
definitions contained in this document with regards to Ginga and NCL.

\subsection{Monomedia support}
The media types in Table~\ref{mediatypes} must be supported by a
``Ginga Full Receiver Profile'' compatible receiver.

\begin{table}[H]
\centering
\caption{``Ginga Full Receiver Profile'' supported media types.}
\label{mediatypes}
\begin{tabularx}{\textwidth}{|p{2.5cm}|X|X|p{2.8cm}|}
  \hline
  Category & Media Type & MIME Type & Extension(s) \\
  \hline
  Image & PNG & image/png & png \\
  \cline{2-4}
   & JPEG & image/jpeg & jpg, jpeg \\
  \cline{2-4}
   & HEIF (HEVC codec) & image/heic & heif, heic \\
  \hline
  MPEG Audio & Supported encoder/profiles are: AAC Profile, Level 4, High 
  Efficiency AAC Profile, Level 4, High Efficiency AAC Profile v2, Level 4, 
  as defined in ISO/IEC 14496-3, and Extended HE AAC Profile, Level 4, as 
  defined in ISO/IEC 23003-3, in both 960 and 1024 transform length. & audio/mp4 & mp4,mpeg4 \\
  \hline
  Vector Graphics & SVG Tiny 1.2 & image/svg+xml & svg,svgz \\
  \hline
  Voice synthesis & SSML 1.1 & application/ssml+xml & ssml \\
  \hline
  Text & Plain text & text/plain & txt \\
  \hline
  Application & ginga-NCL & application/x-ginga-NCL & ncl \\
  \cline{2-4}
   & ginga-NCLua & application/x-ginga-NCLua & lua \\
  \cline{2-4}
   & ginga-HTML5 & text/html & html \\
  \cline{2-4}
   & ncl-settings & application/x-ncl-settings & - \\
  \cline{2-4}
   & ncl-time & application/x-ncl-time & - \\
  \hline
\end{tabularx}
\end{table}

Beyond the media types presented in Table~\ref{mediatypes}, also
Journaline~\cite{ETSI-102979-2008} presentation shall be supported.


\section{Conclusion}
\label{sec:conclusion}

This report aims to contribute to the proper implementation of sound and
multimedia broadcasting in Latin America region. The idea of a harmonized
middleware for both Digital TV and Digital Radio in the region makes sense
as we expect devices like TV sets, cellphones and automotive infotainment
systems to have both Digita TV and Radio (ISDB-T and DRM) tuners in the same
device.
\pagebreak

\addtocontents{toc}{\protect\setcounter{tocdepth}{1}}

\addcontentsline{toc}{section}{References}
\bibliography{references}{}
\bibliographystyle{acm}

\pagebreak
\begin{appendices}

\section{Auxiliary Data Message}
\label{sec:annex}

The MOT protocol provides most of the features required for transporting NCL
applications to be executed by Ginga-ready client receivers.
However, there are three important missing features:
\begin{enumerate}
  \item Transmission and maintenance of independent time
        bases~\cite{Moreno-M-F-2015};
  \item Transport of live editing commands~\cite{soares_nested_2006}; and
  \item Transport of gloss symbols of sign languages, which are used to present
        sign language symbols in the client receiver, for hearing impaired
        users.
\end{enumerate}

The DRM standard already provides means to transmit an absolute clock, but
it does not support independent time bases.
An independent time base is especially important if one wishes to achieve a
fine synchronization between the main audio content and events in the
application, independently from the absolute clock.

Ginga editing commands provide support for the live adaptation of a running
application, such as changing some application structures, start and stop
medias, etc.

Finally, the transport of gloss symbols permits the synthesis of sign
language symbols in the receiver, which is especially important for hearing
impaired users~\cite{araujo_automatic_2012}, and it is also a requirement of
the Brazilian DTV standard~\cite{ABNT-15610-3-2016}.

In DTV, the above data are transmitted using the Stream Events of
DSM-CC~\cite{ISO-13818-6-1998}. However, DSM-CC is not available in Digital
Radio scenario, since it is huge overhead for the available
bandwidth.

In order to address the transmission of such features, new types of messages
must be defined to provide a generic mean to transmit data that
does not fit in the data carousel model of the MOT protocol. Auxiliary Data
Message (ADM) is defined in this subsection for this purpose.

ADM is carried in the same DRM data stream where NCL applications are
carried, but using different \emph{Data group type} on the
\emph{MSC data group}, also used by the MOT protocol, and defined in the DAB
standard~\cite{ETSI-300401-2006}. \emph{Data group type} has a 4 bit size
(0 up to 15).

MOT protocol already uses the 4-bit \emph{Data group type} values 3 (MOT
header), 4 (MOT body), 5 (scrambled MOT body and CA parameters), 6
(uncompressed MOT directory) and 7 (compressed MOT directory). DAB standard
defines \emph{Data group type} values 0 (General data) and 1 (CA messages).

The ADM types use different \emph{Data group type} values. They are listed
in table~\ref{tableadmtypes}.

\begin{table}[H]
\centering
\caption{\emph{Data group type} values for ADM types.}
\label{tableadmtypes}
\begin{tabular}{|l|l|}
  \hline
  \emph{Data group type} value & ADM type \\
  \hline
  10 & TimeBase message \\
  \hline
  11 & EditingCommand message \\
  \hline
  12 & SignLanguage message \\
  \hline
\end{tabular}
\end{table}

The \emph{MSC data group} struture is defined in Chapter 5.3.3 of DAB
standard~\cite{ETSI-300401-2006}. In order to minimize the overhead, as
the \emph{MSC data group} has a variable header length, it's
recommend to use \emph{Extension flag} equal to 0, \emph{Segment flag} set
to 0 and also \emph{User access flag} 0. In order to improve robustness,
CRC flag equal to 1. The other header fiels, which are \emph{Continuity
index} and \emph{Repetition index} must follow the standard, apart from
\emph{Data group type}, which must follow values in
table~\ref{tableadmtypes}. In such configuration, \emph{MSC data group} will
have a 2 byte header in the beginning, and a 2 byte CRC field in the end.
The payload part of \emph{MSC data group}, called \emph{MSC data group data
field} occupies all the rest of the data structure.

Considering the \emph{MSC data group} is mapped directly to a \emph{DRM data
  unit}, and the header size plus CRC size is 4 bytes, and the maximum
payload size carried by a single \emph{MSC data group} is 8191 bytes, an
ADM is limited to 8187 bytes.

\subsection{TimeBase message}

TimeBase messages are used when there is a need the synchronize an audio
content with events in an application.

Table~\ref{tabletb} shows the payload of ADM for the TimeBase type.
A TimeBase message has three fields: \emph{Status},
\emph{DiscontinuityIndicator} and \emph{TimeBaseValue}.

\begin{table}[H]
\centering
\caption{ADM TimeBase syntax.}
\label{tabletb}
\begin{tabular}{|l|l|l|}
  \hline
  TimeBase field & Possible values & Size\\
  \hline
  Status & 0 = Running, 1 = Paused & 1 bit \\
  \hline
  DiscontinuityIndicator & 0 = No, 1 = Yes & 1 bit \\
  \hline
  Reserved for Future Use & - & 5 bits \\
  \hline
  TimeBaseValue & Value & 33 bits \\
  \hline
\end{tabular}
\end{table}

The \emph{Status} field indicates if the time base is running (value equals
to 0) or paused~(value equals to 1).
When a time base is running, the receiver must constantly increment its
current value.
If a time base is running and a message containing a status equal to
\emph{paused} is transmitted, the receiver must stop incrementing the time
base in the indicated \emph{TimeBaseValue} and keep the time base paused.
When the time base is paused and a message containing a status equal to
\emph{Running} is received, the receiver must start increasing the time
base.

The current value of a time value---i.e. the \emph{TimeBaseValue} field---is
a 33 bit value that must be updated at audio super frame cadence (each super
frame sums 1000 to the TimeBaseValue, for improved time resolution).
The \emph{TimeBaseValue} update is driven by the receiving super
frames.
When the last packet that completes a DRM Data Unit containing a TimeBase
message arrives in a given DRM super frame, the \emph{TimeBaseValue}
indicates the moment when the first audio sample, from the first complete
audio frame present in the super frame is played.

A message containing the \emph{DiscontinuityIndicator} set to \emph{Yes}
means that a leap in time has occurred, and no application events must be
triggered when a leap occurs. No time leaps should occur without
\emph{DiscontinuityIndicator} set.

The receiver must compensate internal clock drifts or reception drop-outs
with the values that comes in the ADM TimeBase. If there are small
differences between the internal receiver time base value and the time
base value received in TimeBase messages, the receiver must use elastic
time compensation and not rewind the internal time base value. The only
case where a time base value goes back is when it reaches it greatest
value~\footnote{A broadcaster must avoid time base value to wrap around,
restarting the time base each radio program} and wrap around, or when the
DiscontinuityIndicator is set.

From an NCL application, one must reference a time base value using
a time base value followed by ``tbv'' suffix, like the example
below:

\begin{center}
    <area id="anchor" first="5000tbv"/>
\end{center}

\subsection{EditingCommand message}
Table~\ref{tableed} shows the payload of ADM for the EditingCommand
type.
The syntax of the EditingCommand is composed the field
\emph{EventId}, \emph{DoItNow}, \emph{TimeBaseValue},
\emph{CommandTag} and \emph{CommandPayload}.
\emph{EventId} is a unique event identifier for the command transmission.

\emph{DoItNow} flag indicates if the command should be executed immediately
at arrival in the receiver or not.

\emph{TimeBaseValue} contains the time base value
the receiver should execute the command if the \emph{DoItNow} flag is set to
0.

\emph{CommandTag} indicates the command type and \emph{CommandPayload}
contains the command payload itself. The \emph{CommandPayload} content must
follow the same rules ITU H.761; the only diffence is that the time values should be
interpreted as TBV~(Time Base Value) instead of NPTs~(Normal Play Time).

\begin{table}[H]
\centering
\caption{ADM EditingCommand syntax.}
\label{tableed}
\begin{tabularx}{\textwidth}{|X|X|X|}
  \hline
  EditingCommand field & Possible values & Size\\
  \hline
  EventId & Event Identifier & 16 bits \\
  \hline
  DoItNow & 0 = No, 1 = Yes & 1 bit \\
  \hline
  Reserved for Future Use & - & 6 bits \\
  \hline
  TimeBaseValue & Moment to execute the event & 33 bits \\
  \hline
  CommandTag & The command identifier & 8 bits \\
  \hline
  CommandPayload & The command payload & N bytes \\
  \hline
\end{tabularx}
\end{table}

\subsection{SignLanguage message}
The proposed support for sign language messages is based on the Brazilian DTV
standard ABNT NBR 15610-3~\cite{ABNT-15610-3-2016}, named
LibrasTV~\footnote{Libras is the Brazilian sign language}.

When a radio receiver receives sign language messages in a data stream
associated with the active service, and if the sign language playback option
in enabled in the native receiver software, the sign language must be
synthesized in the screen.

In DRM, the sign language messages are transported in ADM with minor
differences to the DTV standard:

\begin{itemize}
  \item the \emph{eventNPT} fields should be understood as \emph{eventTBV},
        with the time base semantics specified in this text;
  \item the \emph{StreamEventDescriptor} should be used as ADM
        payload, with the same syntax, except for the removal of the
        \emph{descriptorTag} and \emph{descriptorLength} fields;
  \item in the \emph{libras\_content\_type} field of the sign language
        message, the mode 0x01 cannot be used, as video transmission over DRM
        takes too much bitrate;
  \item all the DTV specific fields should be ignored by the receiver.
\end{itemize}

\end{appendices}

\end{document}